\documentclass[a4paper,11pt]{article}
\usepackage{jheppub}
\usepackage[utf8]{inputenc}

\usepackage{lmodern}

\usepackage{amsfonts}
\usepackage{amssymb}
\usepackage{mathrsfs}
\usepackage{mathtools}
\usepackage{mleftright}

\DeclareFontFamily{OT1}{pzc}{}
\DeclareFontShape{OT1}{pzc}{m}{it}{<-> s * [1.200] pzcmi7t}{}
\DeclareMathAlphabet{\mathpzc}{OT1}{pzc}{m}{it}

\def\X{X}
\def\Y{Y}
\def\W{W}
\def\y{z}
\def\aw{\alpha_{\text{w}}}
\def\mflux{\boldsymbol{\mathfrak{m}}}
\def\V{\mathbf{V}}

\def\Xint#1{\mathchoice
    {\XXint\displaystyle\textstyle{#1}}%
    {\XXint\textstyle\scriptstyle{#1}}%
    {\XXint\scriptstyle\scriptscriptstyle{#1}}%
    {\XXint\scriptscriptstyle\scriptscriptstyle{#1}}%
      \!\int}
\def\XXint#1#2#3{{\setbox0=\hbox{$#1{#2#3}{\int}$}
    \vcenter{\hbox{$#2#3$}}\kern-.5\wd0}}
\def\dashint{\Xint-}

\usepackage{tikz}
\usetikzlibrary{
  decorations.markings,
  decorations.pathmorphing,
}
\def\graphwidth{2.0}
\def\graphheight{0.8}
\def\cutval{0.7}
\def\contourgap{0.12}
\def\axescolor{black}
\def\contourcolor{red}
\tikzset{cutstyle/.style={decorate, decoration={zigzag, segment length=6, amplitude=2}, draw=black}}
\tikzset{arrow data/.style 2 args={decoration={markings, mark=at position #1 with \arrow{#2}}, postaction=decorate}}
\tikzset{reverse arrow data/.style 2 args={decoration={markings, mark=at position #1 with {\arrow[xscale=-1]{#2}}}, postaction=decorate}}
\newcommand\drawdot[3]{
    \draw[fill] (#1, 0) circle (.7pt)
    node[black, shift=#2] {#3};
}
\newcommand\laplace{
    \begin{tikzpicture}[scale=2, thick]
    \draw[-latex, \axescolor] (-\graphwidth/3, 0) -- (+\graphwidth, 0) node[right] {$\operatorname{Re}s$};
    \draw[-latex, \axescolor] (0, -\graphheight) -- (0, +\graphheight) node[below left] {$\operatorname{Im}s$};
    \drawdot{0}{(-120:0.5)}{$0$};
    \drawdot{{\cutval*2}}{(-60:0.5)}{$1$};
    \draw[cutstyle] (0, 0) -- ({\cutval*2}, 0);
    \draw[\contourcolor, ultra thick, arrow data={0.3}{latex}, arrow data={0.8}{latex}]
        (0,\contourgap) arc (90:270:\contourgap)
        ({\cutval*2},-\contourgap) -- (0,-\contourgap)
        ({\cutval*2},-\contourgap) arc (-90:90:\contourgap)
        (0,\contourgap) -- ({\cutval*2},\contourgap) node[above left] {$\gamma$};
    \draw[blue,fill=blue]
        ({2*\cutval-0.5}, 0) circle (1pt) node[label={[label distance=2pt]-80:$s_+$}] {}
        (0.5, 0) circle (1pt) node[label={[label distance=2pt]-100:$s_-\!\!$}] {};
    \end{tikzpicture}
}
\newcommand\steepestdescent{
    \begin{tikzpicture}[scale=2, thick]
    \draw[-latex, \axescolor] (-\graphwidth/3, 0) -- (+\graphwidth, 0) node[right] {$\operatorname{Re}s$};
    \draw[-latex, \axescolor] (0, -\graphheight) -- (0, +\graphheight) node[below left] {$\operatorname{Im}s$};
    \drawdot{0}{(-120:0.5)}{$0$};
    \drawdot{{\cutval*2}}{(-60:0.5)}{$1$};
    \draw[cutstyle] (0, 0) -- ({\cutval*2}, 0);
    \draw[\contourcolor, ultra thick, arrow data={0.4}{latex}, reverse arrow data={0.9}{latex}]
        (0,\contourgap) -- ++(0:\graphheight/2.2) arc (-90:-45:\contourgap) -- ++(45:\graphheight) node[below right] {$\gamma'$}
        (0,\contourgap) arc (90:270:\contourgap) -- ++(0:\graphheight/2.2) arc (90:45:\contourgap) -- ++(-45:\graphheight);
    \draw[black!60!green,fill=black!60!green]
        (\cutval, -\graphheight/2) circle (1pt) node[label={[label distance=-3pt]5:$s_-$}] {}
        (\cutval, \graphheight/2) circle (1pt) node[label={[label distance=-3pt]-5:$s_+$}] {};
    \end{tikzpicture}
}

\preprint{UUITP-38/22}

\title{\huge\boldmath The phase diagram of $T\bar{T}$-deformed Yang--Mills theory on the sphere}

\author[a]{Luca Griguolo,}
\author[b]{Rodolfo Panerai,}
\author[a]{Jacopo Papalini,}
\author[c]{and Domenico Seminara}

\affiliation[a]{Dipartimento SMFI, Universit\`a di Parma and INFN Gruppo Collegato di Parma, Viale G.P. Usberti 7/A, 43100 Parma, Italy}
\affiliation[b]{Department of Physics and Astronomy, Uppsala University, Box 516, SE-75120 Uppsala, Sweden}
\affiliation[c]{Dipartimento di Fisica, Universit\`a di Firenze and INFN Sezione di Firenze, via G. Sansone 1, 50019 Sesto Fiorentino, Italy} 

\emailAdd{luca.griguolo@unipr.it}
\emailAdd{rodolfo.panerai@physics.uu.se}
\emailAdd{jacopo.papalini@unipr.it}
\emailAdd{seminara@fi.infn.it}

\abstract{
We study the large-$N$ dynamics of $T\bar{T}$-deformed two-dimensional Yang--Mills theory at genus zero. The 1/$N$-expansion of the free energy is obtained by exploiting the associated flow equation and the complete phase diagram of the theory is derived for both signs of the rescaled deformation parameter $\tau$. We observe a third-order phase transition driven by instanton condensation, which is the deformed version of the familiar Douglas--Kazakov transition separating the weakly-coupled from the strongly-coupled phase. By studying these phases, we compute the deformation of both the perturbative sector and the Gross--Taylor string expansion. Nonperturbative corrections in $\tau$ drive the system into an unexplored disordered phase separated by a novel critical line meeting tangentially the Douglas--Kazakov one at a tricritical point. The associated phase transition is induced by the collision of large-$N$ saddle points, determining its second-order character.
}

\begin{document} 
\maketitle
\flushbottom

\section{Introduction}
Yang--Mills theory in two dimensions is unique due to the absence of local gauge-field excitations as propagating degrees of freedom. The theory is solvable on compact Riemann surfaces of arbitrary topology: its partition function can be exactly computed through different methods, such as lattice techniques \cite{Migdal:1975zg,Rusakov:1990rs,Witten:1991we}, nonabelian localization \cite{Witten:1992xu} or abelianization \cite{Blau:1993hj}. Likewise, observables, such as Wilson-loop correlators, admit nonperturbative evaluation \cite{Rusakov:1990rs}.

Yet, despite its simplicity, the theory retains enough complexity to provide a convenient testing ground for conjectured properties of higher-dimensional models.
Specifically, it can be used as a toy model to study various features of the large-$N$ dynamics of gauge theories, such as the analyticity of the strong coupling expansion \cite{Douglas:1994zu}, or the 't~Hooft gauge/string duality \cite{tHooft:1974pnl}.
In fact, two-dimensional Yang--Mills theory has an exact description at large $N$ in terms of a string theory, with $1/N$ playing the role of the string coupling constant. The expression for the $1/N$-expansion of the free energy can be computed in terms of branched covers of the two-dimensional target space, i.e.\ as string worldsheets of various windings \cite{Gross:1992tu,Gross:1993hu,Gross:1993yt}.

Further, the partition function on genus-zero manifolds exhibits a large-$N$ phase transition in the total area $a$,\footnote{
    The 't Hooft coupling $\lambda$ and the area $a$ form an adimensional coupling $\alpha = \lambda a$, which is the proper coupling of the theory. 
}
going from a strongly-coupled string-like phase for large $a$ to a weakly coupled phase for small $a$. This is a third-order phase transition first observed by Douglas and Kazakov \cite{Douglas:1993iia}. Its physical origin can be understood from the weak-coupling side in terms of instanton condensation \cite{Minahan:1993tp,Gross:1994mr} or as a divergence of the string expansion when seen from the strong-coupling region \cite{Douglas:1993iia,Taylor:1994zm}.

This analysis has been somehow extended to the so-called \emph{generalized two-dimensional Yang--Mills theory} \cite{Douglas:1994pq,Ganor:1994bq}: in two dimensions, an equivalent formulation of the Yang--Mills action takes the form of a BF-theory action with quadratic potential; yet, theories with more general potentials can be considered and solved by applying the localization procedure \cite{Witten:1992xu}.
In some sense, ordinary Yang-Mills theory belongs to a landscape of pure gauge theories, obtained by deforming the familiar action through irrelevant operators constructed from the field strength. The consistency and the solvability of this family are related to the almost-topological character of the seed theory, still preserved by suitable perturbations.

In the last few years, a peculiar deformation of general two-dimensional relativistic quantum field theories has attracted a considerable amount of interest: this is the so-called \emph{$T\bar{T}$-deformation} \cite{Zamolodchikov:2004ce}. It is an irrelevant deformation induced by a particular local operator, quadratic in the stress-energy tensor. The vacuum expectation value of this operator has special properties. The effect of its deformation was studied in \cite{Smirnov:2016lqw,Cavaglia:2016oda}: after compactification on a Euclidean circle of radius $R$, a simple differential equation controls the evolution of the energy spectrum according to the newly-introduced irrelevant coupling $\mu$, also referred to as \emph{deformation parameter}. The solvability of this deformation seems to provide a consistent way to move against the renormalization-group flow and explore unconventional dynamics at ultraviolet fixed points \cite{Datta:2018thy}.

Notably, $T\bar{T}$-deformed theories seem intrinsically related to two-dimensional gravity \cite{Dubovsky:2017cnj,Dubovsky:2018bmo,Ishii:2019uwk}, to random geometries \cite{Cardy:2018sdv}, and can even be reformulated in terms of string theory \cite{Frolov:2019nrr,Callebaut:2019omt,Tolley:2019nmm}. It is, therefore quite natural to study the $T\bar{T}$ deformation in the context of two-dimensional Yang--Mills theory and, in particular, to explore its effect on the large-$N$ limit and its string-theory avatar. Moreover, in the case of conformal field theories, $T\bar{T}$ deformation was observed, inducing a Hagedorn growth of states for large $\mu>0$ \cite{Datta:2018thy} and a complexification of the spectrum for $\mu<0$. It is undoubtedly interesting to understand if and how the deformation modifies the large-$N$ phase diagram.

We initiated a systematic study of $T\bar{T}$-deformed gauge theories,\footnote{
    The $T\bar{T}$ deformation of gauge fields has also been studied in connection with DBI-like theories \cite{Brennan:2019azg,Conti:2022egv}.}
deriving exact results for the abelian case \cite{Griguolo:2022xcj} and for the nonabelian theory on the sphere \cite{Griguolo:2022hek}. Our results reveal a truncation of the spectrum for $\mu>0$ associated with nonanalyticities in the partition function and the appearance of nonperturbative contribution in the deformation parameter for $\mu<0$. It is a challenging task, though, to study the large-$N$ limit of the theory from the exact expressions obtained at finite $N$.

In the present paper, we study the deformed theory on the sphere in the limit where $N$ is large. In taking this limit, one obtains a nontrivial dynamics by keeping finite the 't Hooft coupling $\lambda$ and the dimensionless combination $\tau = \mu \lambda N^2$, which can be regarded as an effective deformation parameter. Having a new coupling $\tau$ opens up a new direction in the phase diagram of the theory, which in the undeformed case was simply the half-line $\alpha > 0$ (i.e., \ $\lambda > 0$). Indeed, studying the full structure of this phase diagram is one of the main goals of this work.

\begin{figure}[htbp]\label{FIG:phase_diagram}
    \centering
    \includegraphics[]{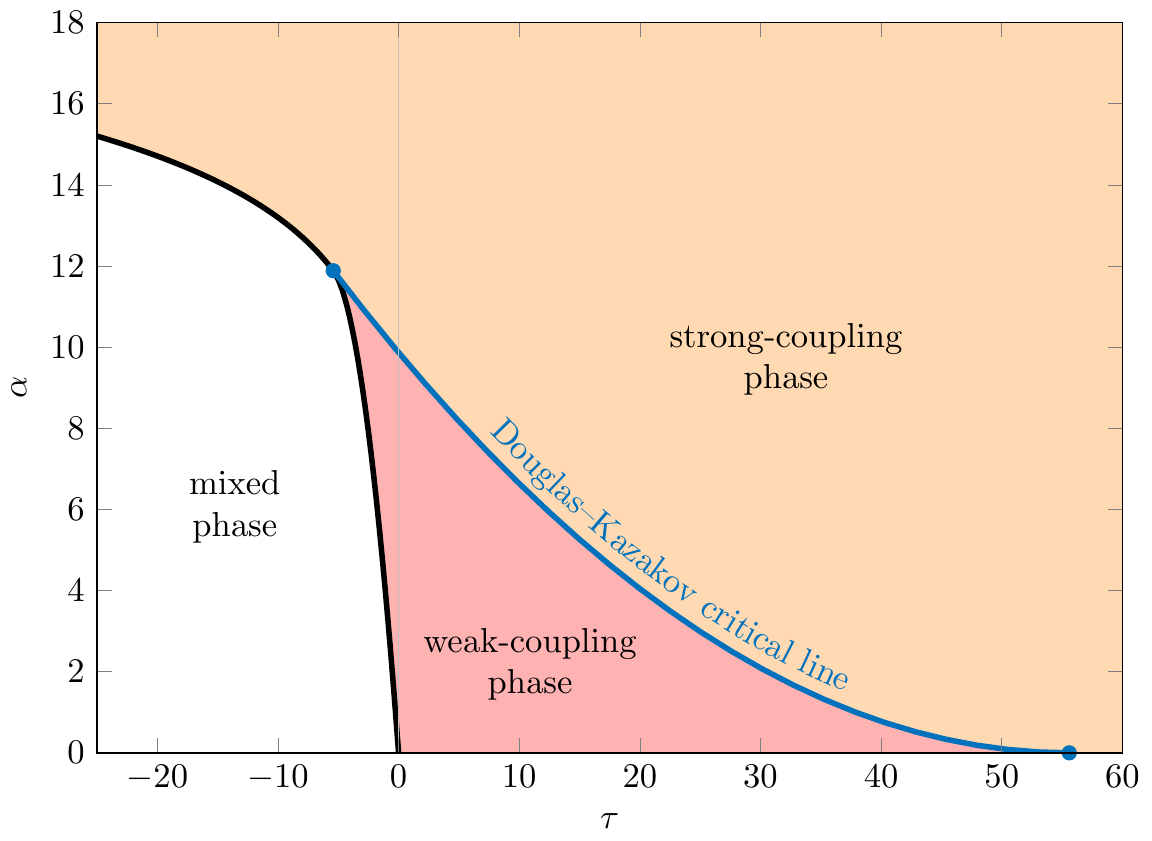}
    \caption{\label{FIG:regions} The phase diagram of the theory at large $N$ has three phases: weak coupling, strong coupling, and mixed coupling. The blue line is the deformed Douglas--Kazakov critical line, associated with a third-order phase transition. The black line is a critical line associated with a second order-phase transition. The two lines join at a multicritical point represented by a blue dot. The thin gray line at $\tau = 0$ corresponds to the undeformed theory.}
\end{figure}

\paragraph{Summary of results.}
Contrary to previous investigations on the subject \cite{Santilli:2018xux,Gorsky:2020qge}, we follow an approach based on iteratively solving the system of partial differential equations governing the deformation of the large-$N$ expansion of the free energy. We find exact solutions at all orders in $1/N$. These are obtained by propagating the initial conditions at $\tau = 0$ associated with both the weak-coupling and the strong-coupling regime along a system of characteristic curves determined by the leading order $F_0$ of the free energy. These curves effectively chart the phase diagram of the deformed theory; much of the information on the large-$N$ dynamics can be obtained by studying their properties. The entire phase diagram is shown in Figure~\ref{FIG:phase_diagram}.

The characteristic curve emanating from the Douglas--Kazakov critical point acts as an interface between the characteristics transporting the weak-coupling and the strong-coupling initial conditions. In other words, the critical point of the undeformed theory is now a critical line with an associated third-order phase transition.
The critical line is monotonically decreasing as a function of $\tau$. It reaches the $\alpha = 0$ axis at a value $\tau_{\text{max}}$ above which the theory exists only in the strong phase. At such a point, the discontinuity of $F_0'''$ diverges.

There is a second endpoint of this critical line where $F_0'''$ again diverges. This happens at $\tau_{\text{mcp}}<0$. We can interpret this behavior by observing that the Douglas--Kazakov line tangentially joins a novel critical line associated with a second-order phase transition on such a point.
This curve is an envelope for the characteristics of both the strong coupling and the weak-coupling phase and effectively acts as a boundary for both phases. From the point of view of the differential equation, this limits the region that can be accessed by propagating the initial condition at $\tau = 0$. This phenomenon has to do with nonperturbative corrections in $\tau$, which cease to be suppressed in the large-$N$ limit upon crossing the envelope. We had already observed in \cite{Griguolo:2022hek} how such corrections introduce ambiguities that can be fixed by imposing a second boundary condition. In the new region, which we refer to as \emph{mixed phase}, the hierarchy between instantons in $\alpha$, typical of the weak-coupling phase, is also lost.

Figure~\ref{FIG:phase_diagram} can then be interpreted as a diagram in which each phase represents a different regime for the instantons. In the weak-coupling phase, all instanton corrections are suppressed. In the strong phase, the instantons in $\alpha$ contribute to the result, while the instantons in $\tau$ are suppressed. Finally, in the mixed phase, both types of instantons contribute.

We mentioned earlier that at finite $N$, the theory exhibits nonanalyticities in the free energy associated with the truncation of the spectrum and with the presence of nonperturbative corrections in the deformation parameter. We can explain the absence of such feature at large $N$ with the way $\tau$ scales with $N$, which makes such a limit well-defined.
For $\tau>0$, the scaling has the effect of restoring an infinite spectrum or, equivalently, of pushing towards $\tau \to +\infty$ the points of nonanalyticity.
On the other hand, if $\alpha>0$, we notice from the phase diagram that there always exists a region for small $\tau<0$ where the nonperturbative corrections in $\tau$ are suppressed,\footnote{
    These corrections have the form $e^{N^2\alpha/2\tau}$ and are thus suppressed at large $N$ for small $\tau$ and large $\alpha$, consistently with the picture emerging from the phase diagram in Figure~\ref{FIG:phase_diagram}.
} thus ensuring the analyticity of the free energy at $\tau = 0$.

\paragraph{Outlook.}
The phase diagram of large-$N$ Yang--Mills theory on the sphere displays an intriguing interplay between different types of nonperturbative contributions. In particular, the discovered second-order phase transition sharply deviates from the familiar third-order Douglas--Kazakov transition, signaling a genuine new effect due to the $T\overline{T}$ deformation. A natural follow-up of the present investigations would consist in deriving an effective description of the mixed phase: in that region of parameters, we expect a behavior dominated by degrees of freedom quite different both from the Gross--Taylor string and from the perturbative gauge fluctuations, typical of the weak-coupling phase.\footnote{
    We already noticed in \cite{Griguolo:2022hek} how the expression for the individual flux sector becomes ill-defined below a certain bound at $\tau_{\text{min}}$.
}

The other obvious extension of our work concerns the study of the large-$N$ Yang--Mills theory on the torus. The undeformed theory has been studied from different points of view over the years. In particular, it admits an accurate string description in the Gross--Taylor approach \cite{Gross:1992tu,Gross:1993hu}, and it is equivalent to a topological string theory on a non-compact toric manifold \cite{Vafa:2004qa}. It would certainly be interesting to understand how these properties are deformed along the $T\overline{T}$ flow and if a string-theory picture survives after the deformation. The torus topology also offers a possible connection with the well-studied case of $T\overline{T}$-deformed conformal field theories: it is well known that Yang--Mills theory on the torus has a large-$N$ description in terms of an interacting compact boson \cite{Douglas:1993wy} with subtle modular properties \cite{Dijkgraaf:1996iy,Okuyama:2019rqn}. We expect that the $T\overline{T}$ deformation could be implemented and studied as some nontrivial interaction potential in this effective theory.

Finally, the large-$N$ theory on the sphere has a dual description in terms of a vicious walkers model \cite{Gorsky:2016mhs}. The Douglas--Kazakov phase transition has been studied in this context \cite{Forrester:2010ah}. It would be nice to extend this duality along the $T\overline{T}$ flow, possibly gaining new understanding of the second-order phase transition.

\section{Yang--Mills theory in two dimensions}\label{SEC:2d_yang_mills}
The partition function for pure Yang--Mills theory in two dimensions on an orientable Riemann surface $\Sigma$ of genus $\mathbf{g}$ and area $a$ can be expressed by the so-called heat-kernel expansion \cite{Migdal:1975zg,Rusakov:1990rs}
\begin{align}\label{EQ:Z_YM}
    Z
    &= \sum_{R} (\dim R)^{2-2\mathbf{g}} \, e^{-g_{\text{YM}}^2a\,C_2(R)/2} \;,
\end{align}
where $g_{\text{YM}}$ is the Yang--Mills coupling.
The sum runs over the equivalency classes $R$ of an irreducible representation of the gauge group $G$, up to isomorphisms. In the above, $C_2(R)$ indicates the eigenvalue of the quadratic Casimir of $R$.

The theory is known to be \emph{almost topological}, i.e.\ its partition function depends on the underlying geometry only through the total area $a$ of the Riemann surface. In this paper, we will study the case of $G\simeq\mathrm{U}(N)$ in the limit where $N$ is large. It is convenient to express the partition function in terms of the effective adimensional coupling $\alpha = \lambda a$, where $\lambda = g_{\text{YM}}^2N$ is the usual 't~Hooft coupling.

In the large-$N$ limit, the theory is conjectured to be dual to some string theory with target space $\Sigma$ \cite{Gross:1992tu,Gross:1993hu,Cordes:1994fc}. Specifically, the Yang--Mills free energy should compute the partition function of a string winding on $\Sigma$ with coupling $g_{\mathrm{s}} = 1/N$ and tension $\lambda$. Evidence for the duality is given by the fact that the $1/N$-expansion of the free energy takes the form
\begin{align}\label{EQ:free_energy}
    F(\alpha)
    &= \log Z(\alpha) \cr
    &= \sum_{\ell=\mathbf{g}}^\infty N^{2-2\ell} \, F_{\ell}(\alpha) \;.
\end{align}
This is consistent with the fact that, according to the Riemann--Hurwitz formula, there are no covering maps between a worldsheet of genus $\ell$ and a two-dimensional target space of genus $\mathbf{g}$, if $\ell<\mathbf{g}$.

\subsection{Genus zero}
We will now focus on the case where $\mathbf{g} = 0$. It is useful to write \eqref{EQ:Z_YM} in a less abstract way by labelling each irreducible representations $R$ of $\mathrm{U}(N)$ through its highest weights $n_1 \geq n_2 \geq \ldots \geq n_N$ in terms of which
\begin{align}
    \dim R
    &= \prod_{i<j}\left(1-\frac{n_i-n_j}{i-j}\right) \;, \\
    C_2(R)
    &= \sum_{i=1}^n n_i(n_i-2i+N+1) \;.
\end{align}
We can then conveniently substitute $n_i = -h_i + i - (N+1)/2$, and find that, in terms of the $h'$s, which now obey $h_1 < h_2 < \ldots < h_N$, we have
\begin{align}\label{EQ:Z_as_sum_over_hs}
    Z(\alpha) = \frac{e^{\alpha(N^2-1)/24}}{G^2(N+1)} \sum_{h_1<\ldots<h_n} \prod_{i<j}(h_j-h_i)^{2} \; e^{-\frac{\alpha}{2N}|\boldsymbol{h}|^2} \;.
\end{align}
The sum runs over integers for odd $N$ and over half-integers for even $N$.

A dual representation of the partition function is easily obtained by using the Poisson summation formula: in doing so, \ref{EQ:Z_as_sum_over_hs} is recast as a sum over unstable instantons \cite{Witten:1992xu,Gross:1994mr}
\begin{align}\label{EQ:Z_from_zhat}
    Z(\alpha) = \sum_{\mflux\in\mathbb{Z}^N} \mathpzc{z}_{\mflux}(\alpha) \;.
\end{align}
The sum runs over the $\mflux$, the Fourier-conjugate of $\boldsymbol{h}$, which can be interpreted as the GNO-quantized magnetic flux vector of a classical solution of the Yang--Mills equation on the sphere. In fact, every term in the sum takes the form
\begin{align}\label{z_m_as_w_times_exp}
    \mathpzc{z}_{\mflux}(\alpha) = w_{\mflux}(\alpha) \; e^{-2\pi^2N|\mflux|^2/\alpha} \;,
\end{align}
where $w_{\mflux}$ is a polynomial capturing the quantum fluctuations about the classical saddle-point action that appears at the exponent. By introducing the differential operator
\begin{align}\label{EQ:differential_operator_V}
    \V = (-4\pi^2)^{-N(N-1)/2} \prod_{i<j} (\partial_{\mathfrak{m}_i} - \partial_{\mathfrak{m}_j})^2 \;,
\end{align}
we can write
\begin{align}\label{EQ:zhat}
    \mathpzc{z}_{\mflux}(\alpha)
    &= \frac{e^{\alpha(N^2-1)/24}}{N!\,G^2(N+1)} \, (-1)^{m} \int_{\mathbb{R}^N} \mathrm{d}h_1 \ldots \mathrm{d}h_N \; \V \, e^{-\frac{\alpha}{2N}|\boldsymbol{h}|^2-2\pi\mathrm{i}\mflux\cdot\boldsymbol{h}} \cr
    &= \mathpzc{z}_{\boldsymbol{0}}(\alpha,0) \, \frac{(\alpha/N)^\nu}{N!\,G(N+1)} \, (-1)^{m} \, \V \, e^{-2\pi^2N|\mflux|^2/\alpha} \;,
\end{align}
where $m = (N-1)(\mathfrak{m}_1+\ldots+\mathfrak{m}_N)$.

The result for the zero-flux and the unit-flux sectors read \cite{Gross:1994mr}
\begin{align}
    \mathpzc{z}_{\boldsymbol{0}}(\alpha) &= C_N \, e^{\alpha(N^2-1)/24} \, \alpha^{-N^2/2} \;, \label{EQ:z_0_undeformed} \\
    \mathpzc{z}_{\boldsymbol{1}}(\alpha) &= (-1)^{N-1} N^{-1}\, e^{-2\pi^2N/\alpha} \, L_{N-1}^1(4\pi^2N/\alpha) \; \mathpzc{z}_{\boldsymbol{0}}(\alpha) \label{EQ:z_1_undeformed} \;,
\end{align}
where we denoted with $\boldsymbol{1}\in\mathbb{Z}^{N}$ a generic unit vector, and we defined
\begin{align}
    C_N = \frac{(2\pi)^{N/2}N^{N^2/2}}{G(N+1)} \;.
\end{align}
The former corresponds to the contribution coming from the vacuum sector and describes the perturbative regime of the theory.
The latter captures the contribution of the first nontrivial solution associated with a monopole configuration of unit flux and classical action $2\pi^2N/\alpha$.

\subsection[The large-\texorpdfstring{$N$}{N} limit]{\boldmath The large-$N$ limit}
At leading order, the large-$N$ limit analysis can be efficiently tackled by approximating the sum in \eqref{EQ:Z_as_sum_over_hs} through the functional integral \cite{Douglas:1993iia}
\begin{align}
    Z &= \int [\mathrm{d}h] \; e^{-N^2S_{\text{eff}}[h]}  \;, \\
    S_{\text{eff}}[h] &= - \frac{\alpha}{24} - \frac{3}{2} + \frac{\alpha}{2}\int_0^1\mathrm{d}x\;h^2(x) - \int_0^1\mathrm{d}x\int_0^1\mathrm{d}y \; \log|h(x)-h(y)| \;,
\end{align}
where the integral is performed over the function $h:[0,1]\to\mathbb{R}$ obeying the constraint $h' \geq 1$. Interestingly, the saddle-point approximation of the above is analogous to that of a Gaussian matrix model. In fact, the density $\rho(h) = \partial x/\partial h$ obeys the saddle-point equation
\begin{align}
    \frac{\alpha}{2} \, h = \dashint \frac{\rho(s)}{h-s} \, \mathrm{d}s \;.
\end{align}
What makes this model nontrivial, however, is the presence of the constraint on $h'$. This implies that a general solution of the above should be of the form
\begin{align}
    \rho(s) = \begin{cases}
        1 & \text{for $|s| < b$} \;, \\
        u(s) & \text{for $b \leq |s| < a$} \;, \\
        0 & \text{for $a \leq |s|$} \;.
    \end{cases}
\end{align}
For $\alpha<\pi^2$, one finds that $b=0$ and $\rho$ obey the typical Wigner semicircle law. For $\alpha > \pi^2$, instead, $b>0$ and to find the density $\rho$ one should solve
\begin{align}
    \frac{\alpha}{2} \, h - \log\frac{h-b}{h+b} = \dashint_{-a}^{-b} \frac{u(s)}{h-s} \, \mathrm{d}u + \dashint_{+b}^{+a} \frac{u(s)}{h-s} \, \mathrm{d}u \;.
\end{align}

The saturation of the constraint on $h'$ is responsible for a third-order phase transition at $\alpha = \pi^2$ that the theory undergoes in the large-$N$ limit, first observed by Douglas and Kazakov \cite{Douglas:1993iia}. Later, in \cite{Gross:1994mr}, it was shown that the transition is induced by instantons. By evaluating the ratio between the unit-flux and the zero-flux partition functions, one can see that for small values of the effective 't Hooft coupling, the former is exponentially suppressed in $N/\alpha$ only for $\alpha<\pi^2$. Specifically, by taking the large-$N$ limit of \eqref{EQ:z_0_undeformed} and \eqref{EQ:z_1_undeformed} below the critical point, one finds
\begin{align}\label{EQ:z_1_z_0_ratio}
    \log\frac{\mathpzc{z}_{\boldsymbol{1}}(\alpha)}{\mathpzc{z}_{\boldsymbol{0}}(\alpha)} \sim -\frac{2\pi^2N}{\alpha}\,\gamma(\alpha/\pi^2) \;,
\end{align}
where
\begin{align}
    \gamma(z) = \sqrt{1-z} - \frac{z}{2} \log\frac{1+\sqrt{1-z}}{1-\sqrt{1-z}} \;.
\end{align}
The function $\gamma(z)$ is positive for $z<1$, i.e.\ in the weak phase, but vanishes as its argument reaches the critical value $z=1$.

The large-$N$ limit of the theory is characterized by the leading order of the free energy in the $1/N$ expansion, which we can write as
\begin{align}\label{EQ:F0_undeformed}
    F_0(\alpha) = \frac{3}{4}+\frac{\alpha}{24}-\frac{\log\alpha}{2} + \Theta(\alpha-\pi^2) \, \Delta F_0(\alpha) \;,
\end{align}
where $\Theta$ denotes the Heaviside step function. The function $\Delta F_0$ captures the behavior above the transition. Its derivative reads \cite{Douglas:1993iia}
\begin{align}\label{EQ:deltaF0_prime_undeformed}
    \partial_\alpha \Delta F_0(\alpha) = \frac{1}{2\alpha} - \frac{4(k(\alpha)+1)\,K^2(k(\alpha))}{3\alpha^2} - \frac{8(k(\alpha)-1)^2\,K^4(k(\alpha))}{3\alpha^3} \;,
\end{align}
where $k(\alpha)$ is obtained by inverting
\begin{align}\label{EQ:m_from_alpha_implicit}
    \alpha = 4K(k)\,(2E(k)+(k-1)\,K(k)) \;.
\end{align}
Here, $K$ and $E$ denote elliptic integrals of the first and second kind, respectively. Near the transition point,
\begin{align}\label{EQ:delta_F_undeformed}
    \Delta F_0(\alpha) = -\frac{(\alpha-\pi^2)^3}{3\pi^6} + O((\alpha-\pi^2)^4) \;,
\end{align}
which shows, indeed, that the transition is of the third order.

For large values of $\alpha$, the free energy is given by the expansion \cite{Douglas:1993iia}
\begin{align}\label{EQ:F0_strong_undeformed}
    F_0(\alpha) = 2e^{-\alpha/2} + (\alpha^2/2-2\alpha-1)e^{-\alpha} + (\alpha^4/3-8\alpha^3/3+4\alpha^2+8/3)e^{-3\alpha/2} + \ldots \;,
\end{align}
that is perfectly consistent with the Gross--Taylor string expansion.

\section{\texorpdfstring{\boldmath $T\bar{T}$}{TT}-deformation}
The flow of the Yang--Mills partition function along the $T\bar{T}$ deformation is controlled by the partial differential equation \cite{Conti:2018jho,Ireland:2019vvj,Santilli:2020qvd,Griguolo:2022hek}
\begin{align}\label{EQ:Z_flow_equation}
    \frac{1}{\lambda}\frac{\partial Z}{\partial \mu} + 2\alpha \, \frac{\partial^2 Z}{\partial \alpha^2} = 0 \;.
\end{align}
For $\mu>0$, we showed in \cite{Griguolo:2022hek} that the deformed partition function is given by a formula analogous to the heath-kernel expansion \eqref{EQ:Z_YM}, namely
\begin{align}\label{EQ:Z_tau_positive}
    Z
    &= \sum_{C_2(R,\mu)>0} (\dim R)^{2} \; e^{-\frac{\alpha\vphantom{C_2}}{2N} \, C_2(R,\mu)} \;,
\end{align}
where each representation is weighted by the ``deformed quadratic Casimir''
\begin{align}
    C_2(R,\mu) &= \frac{C_2(R)}{1-\mu \lambda\,C_2(R)/N} \;,
\end{align}
and the sum is restricted over the representations for which the above is positive. In other words, whenever the deformation parameter reaches a critical value $\mu_R = N/(\lambda\,C_2(R))$, the associated representation $R$ is removed from the sum in \eqref{EQ:Z_tau_positive}. As a consequence, $Z$ is nonanalytic yet smooth for $\mu\in\{\mu_R\}$. Furthermore, for any $\mu>0$, only a finite number of representations $R$ contribute to the partition function, i.e.\ such that $\mu_R<\mu$. The only representation always present in the sum is the trivial representation since it has $C_2=0$. Next, we find that the two $\mathrm{U}(N)$ representations with the smallest Casimir are the fundamental and the antifundamental representation, namely
\begin{align}
    \mathbf{n}_{\text{\makebox[\widthof{A}][c]{F}}} &= (+1,0,\ldots,0) \;, \\
    \mathbf{n}_{\text{A}} &= (0,\ldots,0,-1) \;,
\end{align}
both of which have $C_2=N$. This means that for every $N$, the theory becomes completely trivial when $\mu>1/\lambda$.

A large-$N$ theory with a finite number of states would necessarily bear no resemblance to the two-phases undeformed theory described in the previous section. To obtain a deformed theory with rich dynamics at large-$N$, one should find an appropriate double-scaling limit where $\mu\to0$ when $N\to\infty$, so that the sum over an infinite number of representations is restored. The flow equation \eqref{EQ:Z_flow_equation} suggests the correct scaling. If we consider just the leading order in the large-$N$ expansion of the free energy, namely $\log Z \sim N^2F_0$, the corresponding differential equation reads
\begin{align}
    \frac{1}{\lambda}\frac{\partial F_0}{\partial \mu} + 2N^2\alpha \, \bigg(\frac{\partial F_0}{\partial \alpha}\bigg)^{\!2} + 2\alpha \, \frac{\partial^2 F_0}{\partial \alpha^2} = 0 \;.
\end{align}
By defining as in \cite{Santilli:2020qvd, Griguolo:2022hek} the rescaled adimensional deformation parameter $\tau = \mu\lambda N^2$ we provide the right scaling so that the representations contributing to the leading order of the free energy are still present. At the same time, the deformed Casimir remains nontrivial over such a set when $N$ is large. The flow equation for $F_0$ in terms of $\tau$ then reads
\begin{align}
    \frac{\partial F_0}{\partial \tau} + 2\alpha \, \bigg(\frac{\partial F_0}{\partial \alpha}\bigg)^{\!2} &= 0 \;.
\end{align}

For $\mu<0$, the deformed partition function receives nonperturbative corrections carrying an overall factor of $e^{N^2\alpha/(2\tau)}$, thus making the partition function nonanalytic at $\tau=0$. While we refer the reader to \cite{Griguolo:2022hek} for more detail on the finite-$N$ result, here we notice that at large-$N$, because of the chosen scaling in $N$, one expects these instanton-like corrections to be suppressed for small $\tau$, thus making $F_0$ analytic at $\tau=0$. In the next section, we will see that this is indeed the case.

Let us now quickly review some features of the deformed theory at finite $N$. In \cite{Griguolo:2022hek}, the deformed partition function on the sphere was computed by first finding the correct solution of the flow equation associated with each deformed $\mathpzc{z}_{\mflux}(\alpha,\tau)$, and by then summing over $\mflux$. The partition functions of the various flux sectors are conveniently expressed in terms of the variables
\begin{align}
    \X &= \frac{N^2(N^2-1)\alpha}{2(N^2(12+\tau)-\tau)} \;, \\
    \Y &= \frac{N^2(12+\tau)-\tau}{24\tau} \;, \\
    \W &= \frac{6N^4\alpha}{\tau(N^2(12+\tau)-\tau)} \;,
\end{align}
and read
\begin{align}\label{EQ:z_m_deformed}
    \mathpzc{z}_{\mflux}(\alpha,\tau) =
    \begin{cases}
        \displaystyle
        C_N \, e^{\X} \, \Y^{N^2/2} \sum_{s=0}^{\infty} \frac{p_{\mflux,s}}{s!} \, (-\Y)^s \; U(N^2/2+s,0,\W) & \text{for $\tau>0$} \;, \\[20pt]
        \begin{aligned}[b]
            &{-}\pi C_N \, \W e^{\X} (-\Y)^{N^2/2} \\
            &\times \sum_{s\in K} \frac{(-1)^{2s} \, p_{\mflux,s}}{s!\,\Gamma(s+N^2/2)} \, (-\Y)^s \; {}_1F_1(N^2/2+s+1;2;\W)
        \end{aligned} & \text{for $\tau>0$, $N$ odd} \;,
    \end{cases}
\end{align}
with $K = \{1-\frac{N^2}{2},2-\frac{N^2}{2},\ldots,-\frac{1}{2},0,\frac{1}{2},1,\ldots\}$. For simplicity, we will not deal with the case of even $N$ when $\tau<0$.
The coefficients that appear in the solutions are given by 
\begin{align}\label{EQ:p_s_generic}
    p_{\mflux,s} = 
    \begin{cases}
        \delta_{s,0} & \text{for $\mflux = \boldsymbol{0}$,} \\[10pt]
        \displaystyle
        \frac{(-1)^{m+\nu}N^s}{N!\,G(N+1)} \, \frac{\Gamma(s+1)}{\Gamma(s+1+\nu)} \; \V \, \big(2\pi^2|\mflux|^2\big)^{s+\nu}  & \text{for $\mflux \neq \boldsymbol{0}$,}
    \end{cases}
\end{align}
where $\nu = N(N-1)/2$. This ensures that the limit
\begin{align}\label{EQ:z_m_undeformed_limit}
    \lim_{\tau\to0} \mathpzc{z}_{\mflux}(\alpha,\tau) &= \mathpzc{z}_{\boldsymbol{0}}(\alpha,0) \, \sum_{s=0}^{\infty} \frac{p_{\mflux,s}}{s!} \, (-\alpha)^{-s}
\end{align}
matches the correct expression for the undeformed flux sector \eqref{EQ:zhat}.

\section{The large-\texorpdfstring{\boldmath$N$}{N} expansion of the free energy}\label{SEC:large_N}
In the last section, we have determined the correct scaling of the effective deformation parameter $\tau$, deriving the flow equation that governs the leading order of the free energy in the large-$N$ limit. The goal now is to study further the flow equation, to obtain exact results for all orders in the 1/$N$-expansion, and to identify the main features of the phase diagram of the theory for both positive and negative values of $\tau$.
 
We first need to write down the flow equation acting on the deformed free energy $F(\alpha,\tau) = \log Z(\alpha,\tau)$.
Eq.\ \eqref{EQ:Z_flow_equation} induces a partial differential equation for $F(\alpha,\tau)$ which takes the form
\begin{align}\label{EQ:F_flow_equation}
    N^2\partial_\tau F + 2\alpha (\partial_\alpha F)^2 + 2\alpha\,\partial_\alpha\!\!{}^2F = 0 \;.
\end{align}
Before expanding in powers of $N$, it is useful to transform \eqref{EQ:F_flow_equation} into an equation with constant coefficients by replacing $F$ with
\begin{align}\label{EQ:G_definition}
    F(\alpha,\tau) = N^2G(\sqrt{\alpha},\tau) + \frac{\log\alpha}{4} \;,
\end{align}
thus obtaining for $G(\y,t)$ 
\begin{align}\label{EQ:G_flow_equation}
    \partial_\tau G + \frac{1}{2}(\partial_{\y}G)^2 + \frac{1}{2N^2}\,\partial_{\y}\!\!{}^2G = \frac{3}{8N^4\y^2} \;.
\end{align}
Finally, we assume that $F$, and thus $G$, possess an expansion in powers of $1/N^2$, as in the case of the undeformed theory. In particular, we denote
\begin{align}
    G(\y,\tau) = \sum_{\ell=0}^\infty N^{-2\ell} \, G_\ell(\y,\tau) \;.
\end{align}

Let us now start by considering the leading order at large $N$. Instead of directly dealing with the equation for $G_0$,
\begin{align}\label{EQ:G0_flow_equation}
    \partial_\tau G_0 + \frac{1}{2}(\partial_{\y}G_0)^2 = 0 \;,
\end{align}
it is easier to study the equivalent problem for $\mathcal{E} = \partial_{\y} G_0$, which is described by the well-known inviscid Burgers' equation
\begin{align}\label{EQ:f_equation}
    \partial_\tau \mathcal{E} + \mathcal{E} \, \partial_{\y} \mathcal{E} = 0 \;.
\end{align}
Standard solutions are obtained by studying the characteristics of the differential operator $\mathcal{D} = \partial_\tau + \mathcal{E}\,\partial_{\y}$, i.e.\ the solutions of the ordinary differential equation $\mathrm{d}\y/\mathrm{d}\tau = \mathcal{E}$.
According to \eqref{EQ:f_equation}, $\mathcal{E}$ is constant along the characteristics, which are then given by
\begin{align}\label{EQ:characteristic_line}
    \y = \xi + \tau \, \varphi(\xi) \;,
\end{align}
where $\varphi(\xi) = \mathcal{E}(\xi,0)$ and $\xi$ is some integration constant. The original equation \eqref{EQ:f_equation} is solved by simply inverting \eqref{EQ:characteristic_line}, from which one can write the explicit solution
\begin{align}\label{EQ:f_solution}
    \mathcal{E}(\y,\tau) = \varphi(\xi(\y,\tau)) \;.
\end{align}

It is not difficult at this point to derive the equations for the subleading terms in the large-$N$ expansions:
\begin{align}\label{EQ:G_n_equations}
    \mathcal{D} G_1 &= - \frac{1}{2}\,\partial_{\y}\!\!{}^2G_{0} \;, \vphantom{\sum_{k=1}} \cr
    \mathcal{D} G_2 &= - \frac{1}{2}\,\partial_{\y}\!\!{}^2G_{1} - \frac{1}{2} \, (\partial_{\y}G_1)^2 + \frac{3}{8\y^2} \;, \cr
    \mathcal{D} G_\ell &= - \frac{1}{2}\,\partial_{\y}\!\!{}^2G_{\ell-1} - \frac{1}{2} \sum_{k=1}^{\ell-1}\partial_{\y}G_k\,\partial_{\y}G_{\ell-k} \;, \qquad \text{for $\ell\geq2$.}
\end{align}
This recursive system can be conveniently integrated by changing variables with
\begin{align}
    \tilde{G}_\ell(\xi, \tau) = G_\ell(\y(\xi,\tau),\tau) \;,
\end{align}
in terms of which \eqref{EQ:G_n_equations} becomes
\begin{align}
    \partial_\tau \tilde{G}_\ell
    &= \frac{\tau\,\ddot{\varphi}\,\partial_\xi \tilde{G}_{\ell-1}}{2(1+\tau\,\dot{\varphi})^3} - \frac{\partial_\xi\!\!{}^2\tilde{G}_{\ell-1} + S_\ell}{2(1+\tau\,\dot{\varphi})^2} + \frac{3\delta_{\ell,2}}{8\y^2} \;,
\end{align}
where $S_1 = 0$, while for $\ell\geq2$,
\begin{align}
    S_\ell(\xi,\tau) = \sum_{k=1}^{\ell-1}\partial_\xi\tilde{G}_k(\xi,\tau)\,\partial_\xi\tilde{G}_{\ell-k}(\xi,\tau) \;.
\end{align}
The solutions are now easy to find:
\begin{align}
    \tilde{G}_0(\xi,\tau) &= -\frac{1}{2}\int_0^\tau\mathrm{d}t \; \mathcal{E}^2(\y(\xi,\tau),t) + F_0(\y^2(\xi,\tau),0) \;, \label{EQ:G_0_solution} \\
    \tilde{G}_\ell(\xi,\tau)
    &= \int_{0}^\tau \mathrm{d}t \, \bigg( \,\frac{t\,\ddot{\varphi}(\xi)\,\partial_\xi \tilde{G}_{\ell-1}(\xi,t)}{2(1+t\,\dot{\varphi}(\xi))^3} - \frac{\partial_\xi\!\!{}^2\tilde{G}_{\ell-1}(\xi,t)+S_\ell(\xi,t)}{2(1+t\,\dot{\varphi}(\xi))^2} \, \bigg) \cr
    &\quad - \frac{\log\xi \, \delta_{\ell,1}}{2} + \frac{3\tau \, \delta_{\ell,2}}{8\xi(\xi+\tau\,\varphi(\xi))} + F_\ell(\xi^2,0) \;, \qquad \text{for $\ell\geq1$.} \label{EQ:G_n_solution}
\end{align}
In the second identity, we made use of
\begin{align}
    F_\ell(\y^2(\xi,\tau),\tau) &= \tilde{G}_\ell(\xi,\tau) + \frac{\delta_{\ell,1}}{2}\,\log{(\xi+\tau\,\varphi(\xi))} \;,
\end{align}
which is a trivial consequence of \eqref{EQ:G_definition}.
Conversely, we can recover the free energy from the solutions \eqref{EQ:G_0_solution} and \eqref{EQ:G_n_solution} with
\begin{align}
    F_\ell(\alpha,\tau) &= \tilde{G}_\ell(\xi(\sqrt{\alpha},\tau),\tau) + \delta_{\ell,1}\,\frac{\log{\alpha}}{4} \;.
\end{align}

\subsection{The phase diagram}
As a direct application of the previous formulas, one can read off the large-$N$ expansion of the free energy in the weak-coupling phase,\footnote{
    We have neglected inessential constant terms contributing to subleading orders in $1/N$.}
taking as boundary condition the undeformed zero-instanton partition function \eqref{EQ:z_0_undeformed} 
\begin{align}
    F_0(\alpha,0) &= \frac{3}{4}+\frac{\alpha}{24}-\frac{\log\alpha}{2} \;, \label{EQ:F_0} \\
    F_1(\alpha,0) &= -\frac{\alpha}{24} \;, \\
    F_\ell(\alpha,0) &= 0 \;, \qquad\qquad \text{for $\ell\geq2$.} \vphantom{\frac{\alpha}{}}
\end{align}
A peculiar feature of the undeformed theory is that at weak coupling, only $F_0$ and $F_1$ are nontrivial. As we will see in a moment, this property ceases to hold at finite $\tau$.

To show this, we simply apply the algorithm previously described. The first step is to use \eqref{EQ:F_0} to compute
\begin{align}
    \varphi(\xi) = \mathcal{E}(\xi,0) = \frac{\xi}{12} - \frac{1}{\xi} \;.
\end{align}
We then plug this in \eqref{EQ:characteristic_line} and find
\begin{align}
    \xi = 6 \, \frac{\y+\sqrt{\y^2-\aw}}{\tau+12} \;,
\end{align}
which, in turn, from \eqref{EQ:f_solution} and defining  $\aw = -\tau(12+\tau)/3$, gives
\begin{align}\label{EQ:f_weak_coupling}
    \mathcal{E}(\y,\tau) = \frac{\y}{\tau+12} - 2\,\frac{\y-\sqrt{\y^2-\aw}}{\aw} \;.
\end{align}

Before computing the deformed large-$N$ expansion of the free energy, we should discuss the bounds on the validity of the solution \eqref{EQ:f_weak_coupling}. A first bound comes from the fact that the initial condition we imposed so far holds in the weak-coupling phase, i.e.\ when $\alpha<\pi^2$ for the undeformed theory. Therefore, this initial condition can only be propagated in the region of parameters covered by characteristics that cross the $\tau = 0$ axis in the interval $\alpha\in(0,\pi^2)$. In other words, the characteristic
\begin{align}\label{EQ:bound_weak_coupling}
    \alpha = \left[\pi + \tau \left(\frac{\pi}{12} - \frac{1}{\pi}\right)\right]^2
\end{align}
represents a bound for the validity of the solution of the Burgers' equation \eqref{EQ:f_equation} at weak coupling. We see that for $\tau>\tau_{\text{max}}$, where
\begin{align}
    \tau_{\text{max}} = \frac{12\pi^2}{12-\pi^2} \;,
\end{align}
the theory is always in the strong-coupling phase for any value of $\alpha$.

Furthermore, we notice that $\xi$ and, as a consequence, $\mathcal{E}$ are real for $\alpha\geq \aw$. The set of points where the last inequality saturates is the envelope of the system of characteristics \eqref{EQ:f_weak_coupling}. This means that the parabola
\begin{align}\label{EQ:bound_no_tau_instantons}
    \alpha-\aw = 0
\end{align}
represents another bound for \eqref{EQ:f_equation} at weak coupling. In the next section, we will see what the origin of said bound is and how to make sense of the deformed Yang--Mills partition function beyond the envelope.

\begin{figure}[tbp]
    \centering
    \includegraphics[]{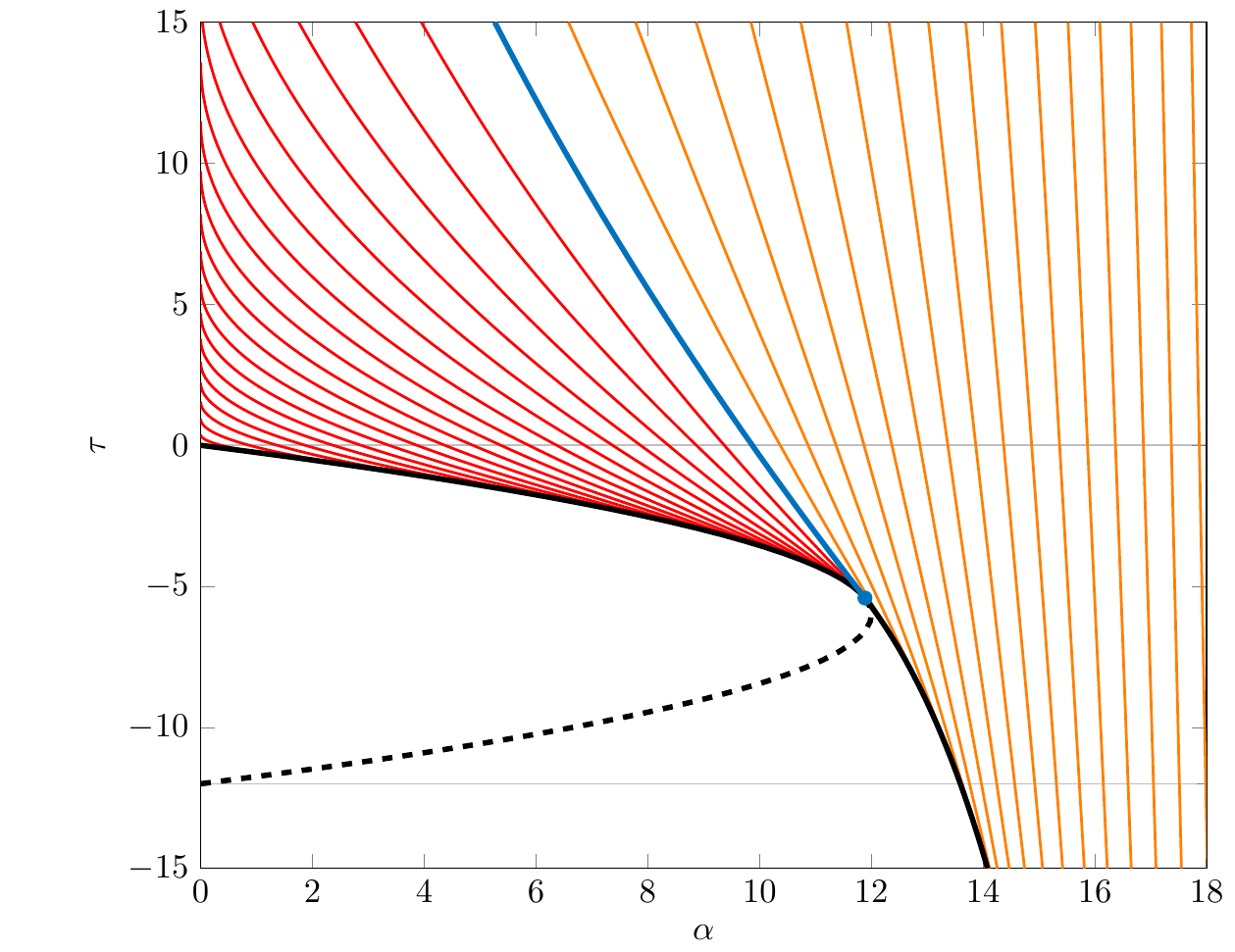}
    \caption{\label{FIG:characteristics} The diagram shows the system of characteristics associated with both the weak-coupling phase (red lines) and the strong-coupling phase (orange lines). The blue line is the characteristic that acts as a critical line between the two phases and crosses the $\tau=0$ axis at $\alpha=\pi^2$. The critical line ends on the multicritical point $(\alpha_{\text{mcp}},\tau_{\text{mcp}})$. The black parabola delimiting the weak-coupling phase is the envelope of the weak-coupling characteristics and corresponds to $\alpha = \aw$. The black line delimiting the strong-coupling phase is the envelope of the strong-coupling characteristics and has coordinates $(\alpha_{\text{s}},\tau_{\text{s}})$.}
\end{figure}

As can be seen in Figure~\ref{FIG:characteristics}, the two parabolas \eqref{EQ:bound_weak_coupling} and \eqref{EQ:bound_no_tau_instantons} are tangent at the multicritical point
\begin{align}\label{EQ:critical_point}
    \alpha_{\text{mcp}} &= \left(\frac{24\pi}{12+\pi^2}\right)^2 \;, & \tau_{\text{mcp}} &= -\frac{12\pi^2}{12+\pi^2} \;.
\end{align}

We can now apply \eqref{EQ:G_0_solution} and \eqref{EQ:G_n_solution} to recursively generate any term in the large-$N$ expansion of the deformed free energy. The first few terms read
\begin{align}
    \label{EQ:F_0_deformed_weak}
    F_0(\alpha,\tau)
    &= \frac{3}{4}+\frac{\tilde{\alpha}}{24}-\frac{\log\tilde{\alpha}}{2} + \frac{\tau(12-\tilde{\alpha})^2}{288\tilde{\alpha}\vphantom{)}} \;, \vphantom{\Bigg(}\\
    F_1(\alpha,\tau)
    &= -\frac{\tilde{\alpha}}{24} - \frac{1}{4}\log\mleft(1 - \frac{\aw}{\alpha}\mright) \;, \\
    F_2(\alpha,\tau)
    &= \frac{\tau^2}{72} \Bigg(\frac{\tilde{\alpha}^2}{\tilde{\alpha}\aw-4\tau^2} + \frac{12\tilde{\alpha}\tau(5\tau-36)}{(\tilde{\alpha}\aw-4\tau^2)^2} + \frac{48\tau^2(21\tilde{\alpha}\aw-4\tau^2)}{(\tilde{\alpha}\aw-4\tau^2)^3} - \frac{108}{\tilde{\alpha}\aw+4\tau^2}\Bigg) \;,
\end{align}
where
\begin{align}
    \tilde{\alpha} &= \alpha \, \frac{1 - \aw/2\alpha + \sqrt{1 - \aw/\alpha}}{2(1+\tau/12)^2} \;.
\end{align}

Notice that the critical line $\tilde{\alpha} = \pi^2$ is not only a characteristic for the weak phase but also for the strong phase. This is simply due to the fact that $\mathcal{E}(\y,0)$ is a continuous function at $\y=\pi$, since the transition is of the third order. Therefore, the line of equation
\begin{align}
    \y = \pi + \tau \varphi(\pi)
\end{align}
is a characteristic shared by both phases. We will discuss the transition in more detail in the next section.

We start now to explore the deformation of the original strong-coupling phase. The relevant characteristics can be obtained in implicit form from \eqref{EQ:F0_undeformed}, \eqref{EQ:deltaF0_prime_undeformed} and \eqref{EQ:m_from_alpha_implicit}
\begin{align}
    \varphi (\xi)= \frac{\xi}{12} - \frac{8(k+1)K^2}{3\xi^3} - \frac{16(k+1)^2K^4}{3\xi^5} \;,
\end{align}
where
\begin{align}\label{EQ:xi_with_k}
    \xi^2 = 4K(2E+(k-1)\,K) \;.
\end{align}
The corresponding curves are plotted in Figure~\ref{FIG:characteristics}. We see that, as it happens for the weak-coupling phase, 
the strong-coupling characteristics have an envelope for some range of negative values of $\tau$. To find such a curve, we need to solve $1+\tau\dot{\varphi}(\xi) = 0$ on the solutions of the characteristics equation \eqref{EQ:characteristic_line}. This is easily done in parametric form. In fact, while it is not possible to invert \eqref{EQ:xi_with_k} in closed form, one can still use it to obtain $k'(\xi)$ as a function of $k$ itself obtaining
\begin{align}\label{EQ:strong_envelope}
    \alpha_{\text{s}} &= \left(\frac{384(k-1)^2K^4\xi}{\xi^6-32(k+1)K^2\xi^2+320(k-1)^2K^4}\right)^{\!2} \;, \cr
    \tau_{\text{s}} &= -\frac{12\xi^6}{\xi^6-32(k+1)K^2\xi^2+320(k-1)^2K^4} \;,
\end{align} 
in terms of the parameter $k\in[0,1)$. The range corresponds to $\xi\in[\pi,\infty)$. The envelope has one extremum, namely the point at $k=0$, that coincides with the multicritical point \eqref{EQ:critical_point}: it connects nicely with the envelope for the weak-coupling characteristics, $\alpha=\alpha_{\mathrm{c}}$, thus creating a continuous line. It is also tangent to both the envelope for the weak-coupling characteristics and the critical line for the Douglas--Kazakov phase transition, as the three curves all share the same derivative at the multicritical point
\begin{align}
    \frac{\mathrm{d}\alpha}{\mathrm{d}\tau} \bigg|_{\text{mcp}} = 4 - \frac{96}{12+\pi^2}.
\end{align}

At large $\alpha$, as already remarked in \eqref{EQ:F0_strong_undeformed}, the leading order of the undeformed free energy scales exponentially as $F_0(\alpha,0) \sim 2e^{-\alpha/2} \;.$ Because
\begin{align}
    \mathcal{E}(\y,0) = -\y\,G_0(\y,0) \sim -2\y\,e^{-\y^2/2}
\end{align}
is exponentially suppressed in $\y$, the characteristic equation \eqref{EQ:characteristic_line} reads $\y \sim \xi$ for large $\y$.
In other words, as $\y$ increases, the $\mathcal{E}$ dependence on $\tau$ gets weaker and weaker, and the characteristics become, essentially, vertical lines in Figure~\ref{FIG:characteristics}.

In this regime, it is convenient to solve the characteristic equation by expanding $\xi$ as a power series in $\tau$ and then by fixing the coefficients of the expansion order by order. The solution,
\begin{align}
    \xi = {}
    &\,\y + 2\y\tau \, e^{-\y^2/2} + (\y^5\tau - 4\y^3\tau^2 - 6\y^3\tau + 4\y\tau^2 + 2\y\tau) \, e^{-\y^2} \cr
    &+ (\y^9\tau - 6\y^7\tau^2 - 32\y^7\tau/3 + 12\y^5\tau^3 + 48\y^5\tau^2 + 28\y^5\tau - 28\y^3\tau^3 - 60\y^3\tau^2 \cr
    &\quad\; - 16\y^3\tau + 8\y\tau^3 + 8\y\tau^2 + 8\y\tau) \, e^{-3\y^2/2} + \ldots \;,
\end{align}
gives, in turn, a power-series expression for $\mathcal{E}(\y,\tau)$. Upon integration, we find
\begin{align}
    F_0(\alpha,\tau) = {}
    &\,2e^{-\alpha/2} + (\alpha^2/2-2\alpha-1-2\alpha\tau)\,e^{-\alpha} \cr
    &+ (\alpha^4/3 - 8\alpha^3/3 + 4\alpha^2 + 8/3 - 2\alpha^3\tau + 12\alpha^2\tau - 4\alpha\tau + 4\alpha^2\tau^2 - 4\alpha\tau^2)\,e^{-3\alpha/2} \cr
    &+ \ldots \;,
\end{align}
which is the $\tau$-deformed version of \eqref{EQ:F0_strong_undeformed}. Some comments are now in order to interpret the above result. The undeformed expression captures the Gross--Taylor string theory on the genus-zero target space \cite{Gross:1993yt}, the leading order corresponding to connected covering maps of the type $S^2 \to S^2$. The exponential terms of the form $e^{-n \alpha/2}$ represent the contributions of coverings of degree $n$, while the associated polynomials are obtained by integrating over the positions of various types of singularities \cite{Gross:1993hu,Gross:1993yt}.
In this regime, the $\tau$ deformation affects the polynomial part and acts as a perturbation of the original string expansion: one could conjecture that the deformation provides a refinement for the maps contributing to the string theory, similarly to the generalization induced by higher Casimirs \cite{Ganor:1994bq,Cordes:1994fc}, but a precise interpretation of the new terms and their geometrical meaning are beyond the scope of the present paper.

\section{The deformed Douglas--Kazakov phase transition}
In the last section, we have seen that the deformed theory exhibits the critical line \eqref{EQ:bound_weak_coupling}, separating the weak-coupling phase from the strong-coupling phase, which is the continuation of the Douglas--Kazakov critical point of the undeformed theory.\footnote{
    In \cite{Santilli:2018xux}, the same phase transition was studied by considering the matrix-model of \cite{Douglas:1993iia} with a $\tau$-deformed potential. 
} The associated phase transition remains of the third order. In fact, from \eqref{EQ:delta_F_undeformed} we see that near the critical line
\begin{align}
    \Delta \mathcal{E}(\y,0)
    &= 2\y \, \partial_\alpha \Delta F_0(\alpha,0)\big|_{\alpha=\y^2} \\
    &= -\frac{\y}{\pi^6} \, (\y^2-\pi^2)^2 + O((\y^2-\pi^2)^3) \;.
\end{align}
Let us consider the second derivative
\begin{align}
    \partial^2_{\y}\Delta \mathcal{E} = \Delta\ddot{\varphi}(\xi) \, (\partial^{\vphantom{2}}_{\y}\xi)^2 + \Delta\dot{\varphi}(\xi) \, \partial^2_{\y}\xi \;,
\end{align}
and evaluate it on the characteristic with $\xi = \pi$. The second term vanishes since $\Delta\dot{\varphi}(\pi) = 0$, and we are left with
\begin{align}
    \partial^2_{\y}\Delta \mathcal{E}(\y(\pi,\tau))
    &= \Delta\ddot{\varphi}(\pi) \, (1+\tau\dot{\varphi}(\pi))^{-2} \cr
    &= -\frac{16}{\pi^3}\bigg(\frac{\tau_{\text{mcp}}}{\tau-\tau_{\text{mcp}}}\bigg)^{\!2} \;.
\end{align}
The discontinuity of the third derivative of the free energy on the critical line is easily obtained as
\begin{align}
    \operatorname{Disc} \partial^3_\alpha F_0 = \frac{2}{\pi^6}\bigg(\frac{\tau_{\text{mcp}}}{\tau-\tau_{\text{mcp}}}\bigg)^{\!2}\bigg(\frac{\tau_{\text{max}}}{\tau-\tau_{\text{max}}}\bigg)^{\!3} \;,
\end{align}
which generalizes the undeformed result in \eqref{EQ:delta_F_undeformed}.
This expression diverges at both $\tau_{\text{mcp}}$ and $\tau_{\text{max}}$, i.e.\ as one approaches both the multicritical point \eqref{EQ:critical_point} and the limit value after which the theory is in the strong phase for any $\alpha$.

As mentioned in Section~\ref{SEC:2d_yang_mills}, the Douglas--Kazakov phase transition of the undeformed theory is driven by instantons, and this fact was argued in \cite{Gross:1994mr} by computing the ratio \eqref{EQ:z_1_z_0_ratio}. We now want to show that this property still holds in the deformed theory.

The first step is to obtain a convenient representation for the instanton contributions, suitable to compute the large-$N$ limit of the relevant ratio. We found useful to express the Tricomi confluent hypergeometric through the following integral representation, that holds for $\operatorname{Re}a>0$ and $\operatorname{Re}z>0$,
\begin{align}\label{EQ:Tricomi_integral_representation}
    U(a,b,z) = \frac{1}{\Gamma(a)} \int_0^{\infty} \mathrm{d}t \; e^{-zt} \, t^{a-1} \, (t+1)^{b-a-1} \;.
\end{align}
The instanton partition function \eqref{EQ:z_m_deformed} can be recast as
\begin{align}
    \mathpzc{z}_{\mflux}(\alpha,\tau) = C_N \, e^{\X} \sum_{s=0}^{\infty} \frac{(-1)^s\,p_{\mflux,s}}{s!\,\Gamma(N^2/2+s)} \int_0^{\infty} \mathrm{d}t \; \frac{e^{-t\W}}{t(t+1)}\bigg(\frac{t\Y}{t+1}\bigg)^{\!N^2/2+s}.
\end{align}
To evaluate its large-$N$ limit, we use the Stirling approximation
\begin{align}
    \frac{C_N}{\Gamma(N^2/2+s)} \sim e^{\frac{5}{4}N^2} \bigg(\frac{2}{N^2}\bigg)^{\!N^2/2+s} \;,
\end{align}
and define
\begin{align}
    \X &\sim N^2x \;, & x&=\frac{\alpha}{2(12+\tau)} \;, \\
    \Y &\sim N^2y \;, & y&=\frac{12+\tau}{24\tau\vphantom{()}} \;, \\
    \W &\sim N^2w \;, & w&=\frac{6\alpha}{\tau(12+\tau)} \;.
\end{align}
We then change the integration variable with
\begin{align}
    \frac{1}{u} = \frac{2ty}{t+1}
\end{align}
and by using \eqref{EQ:z_m_undeformed_limit} we obtain
\begin{align}\label{z_m_large_N}
    \mathpzc{z}_{\mflux}(\alpha,\tau)
    &\sim \int_{\frac{1}{2y}}^{\infty} \frac{\mathrm{d}u}{u} \; e^{-N^2\rho(u)} \; \frac{\mathpzc{z}_{\mflux}(u,0)}{\mathpzc{z}_{\boldsymbol{0}}(u,0)} \;,
\end{align}
where
\begin{align}
    \rho(u) = \frac{w}{2uy-1}-x-\frac{5}{4}+\frac{1}{2}\log u \;.
\end{align}
The function $\rho(u)$ has minimum in $u = \tilde{\alpha}$. Moreover, this saddle point always falls within the integration range since $\tilde{\alpha}>1/(2y)$ for $\tau\geq0$.

As expected, the zero-flux sector \eqref{z_m_large_N} reproduces the result of the large-$N$ leading order at weak coupling computed in \eqref{EQ:F_0_deformed_weak}. Namely, $\rho(\tilde{\alpha}) = -F_0(\alpha,\tau)$, so that
\begin{align}
    \mathpzc{z}_{\boldsymbol{0}}(\alpha,\tau) \sim e^{N^2F_0(\alpha,\tau)} \;.
\end{align}
For a generic $\mflux$, we assume that in the large-$N$ limit, the sum is always subleading with respect to the exponential, i.e. that the sum does not contribute to the fluctuations about the saddle. Under this assumption, one finds
\begin{align}
    \frac{\mathpzc{z}_{\mflux}(\alpha,\tau)}{\mathpzc{z}_{\boldsymbol{0}}(\alpha,\tau)} \sim \frac{\mathpzc{z}_{\mflux}(\tilde{\alpha},0)}{\mathpzc{z}_{\boldsymbol{0}}(\tilde{\alpha},0)} \;.
\end{align}
In other words, the deformed ratio coincides with the undeformed one upon replacing $\alpha$ with $\tilde{\alpha}$.\footnote{
    An analogous result for $\mflux = \boldsymbol{1}$ was obtained in \cite{Santilli:2018xux} through a different approach.
}

Let us see how this works concretely in the case when $\mflux = \boldsymbol{1}$. From \eqref{EQ:p_s_generic}, one can compute the coefficients for the one-flux sector, which turns out to be
\begin{align}\label{EQ:p_s_one_instanton}
    p_{\boldsymbol{1},s}
    &= (-1)^{N-1} \, (2\pi^2N)^s \; {}_2F_1(-s,1-N;2;2) \;.
\end{align}
Then, the associated sum can be performed exactly by using the identity
\begin{align}
    \sum_{s=0}^\infty \frac{\xi^s}{s!} \, {}_2F_1(-s,a;b;x)
    &= e^\xi \, {}_1F_1(a;b;-\xi x) \cr
    &= e^\xi \, \frac{\Gamma(1-a)\Gamma(b)}{\Gamma(b-a)} \, L_{-a}^{b-1}(-\xi x) \;,
\end{align}
and as a result, we find
\begin{align}
    \frac{\mathpzc{z}_{\boldsymbol{1}}}{\mathpzc{z}_{\boldsymbol{0}}} \sim (-1)^{N-1} \, e^{-2\pi^2N/\tilde{\alpha}} \; L_{N-1}^{1}(4\pi^2N/\tilde{\alpha}) \;,
\end{align}
which is the expected result. Therefore, we can simply invoke the argument of \cite{Gross:1994mr} and conclude that the partition function of the unit-flux sector is no longer suppressed in the large-$N$ limit for $\tilde{\alpha} \geq \pi^2$. This confirms the observation of Section \ref{SEC:large_N}, where we obtained the same condition for the transition to the strong-coupling phase of the deformed theory. We remark that the above picture is a smooth deformation of the undeformed case. The nonperturbative contributions driving the transition are still instantons labeled by the quantized magnetic flux vector.

\section{Envelopes and nonperturbative corrections}
A puzzling feature of the phase diagram in Figure~\ref{FIG:characteristics} is the emergence of an envelope of characteristics in both the weak-coupling and the strong-coupling phase. This phenomenon is similar to the emergence of a Douglas--Kazakov phase transition in that instantons drive both. However, while the Douglas--Kazakov transition is due to instantons in the effective 't Hooft coupling $\alpha$, the novel phase transition is due to instantons in the deformation parameter $\tau$.

To show this, let us first focus on the envelope at $\alpha = \aw$. In this case, the analysis is more straightforward since the envelope sits at the boundary of the weak-coupling phase where the zero-flux sector completely dominates the dynamics. We will show that, when $N$ is large, the nonperturbative corrections in $\tau$, typical of the deformation with $\tau<0$, are suppressed only for $\alpha>\aw$.

\begin{figure}[tbp]
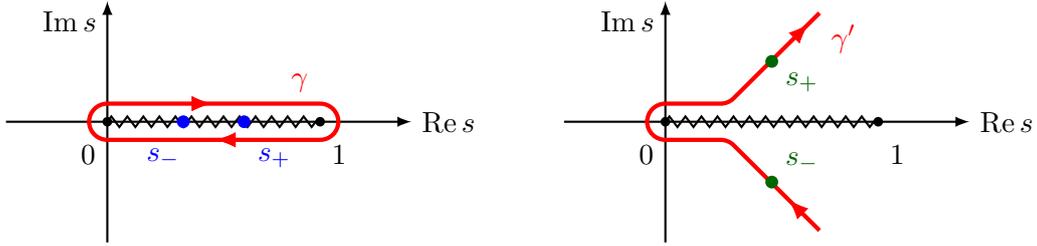

    \centering\laplace\qquad\steepestdescent
    \caption{\label{FIG:contours} On the left, the integration contour for the original integral in \eqref{EQ:Z_tau_neg_integral_representation}. This choice is particularly convenient for $\alpha>\aw$, where both saddles are real (blue dots). On the right, the deformed contour is associated with the steepest descent approximation for $\alpha < \aw$, where both saddles are complex (green dots).}
\end{figure}

Again, we use an integral representation for the Kummer confluent hypergeometric function to conveniently express our zero-instanton partition function. For $\operatorname{Re}a>0$,
\begin{align}\label{EQ:Kummer_integral_representation}
    {}_1F_1(a;b;z) = \frac{1}{2\pi\mathrm{i}} \frac{\Gamma(b)\,\Gamma(a-b+1)}{\Gamma(a)} \int_0^{(1+)} \mathrm{d}t \; e^{zt} \, t^{a-1} \, (t-1)^{b-a-1} \;,
\end{align}
where the integral is taken over a contour starting and ending in 0 and encircling 1 in the positive sense.
Armed with the above representation, we recast $\mathpzc{z}_{\boldsymbol{0}}(\alpha,\tau)$ at finite $N$ and $\tau<0$ in terms of a contour integral and study its large-$N$ limit using a saddle-point approximation. Starting from \eqref{EQ:z_m_deformed} and \eqref{EQ:p_s_generic}, we find
\begin{align}\label{EQ:Z_tau_neg_integral_representation}
    \mathpzc{z}_{\boldsymbol{0}}(\alpha,\tau)
    &= -\frac{\mathrm{i}C_N}{2\,\Gamma(N^2/2+1)} \, \W e^\X (-\Y)^{N^2/2} \oint_\gamma \mathrm{d}s \; e^{\W s} \left(\frac{s}{s-1}\right)^{N^2/2} \;.
\end{align}
The choice of contour $\gamma$ is shown in Figure~\ref{FIG:contours}. When $N$ is large, we can write\footnote{For simplicity, we discard an irrelevant overall constant.}
\begin{align}
    \mathpzc{z}_{\boldsymbol{0}}(\alpha,\tau)
    &\sim
    \frac{\mathrm{i}Ne^{\frac{5}{4}N^2}}{\sqrt{\pi}}
    \, \frac{\alpha}{\aw} \, e^{-N^2\frac{\tau}{6}\frac{\alpha}{\aw}} \left(-\frac{12+\tau}{12\tau}\right)^{N^2/2} \oint_\gamma \mathrm{d}s \; e^{-N^2\phi(s)} \;,
\end{align}
where
\begin{align}
    \phi(s) = 2\,\frac{\alpha}{\aw}\,s-\frac{1}{2}\log\frac{s}{s-1} \;.
\end{align}
As $N\to\infty$ the integral will be dominated by the stationary points of $\phi(s)$,
\begin{align}
    s_{\pm} &= 
    \begin{cases}
        \displaystyle\frac{1}{2}\left(1\pm\phantom{\mathrm{i}}\sqrt{1-\frac{\aw}{\alpha}}\right) \qquad \text{for $\alpha>\aw$ ,} \\[1em]
        \displaystyle\frac{1}{2}\left(1\pm\mathrm{i}\sqrt{\frac{\aw}{\alpha}-1}\right) \qquad \text{for $\alpha<\aw$ .}
    \end{cases}
\end{align}
For $\tau = 0$, the stationary points are at the endpoints $s=0$ and $s=1$ of the branch cut of $\phi(s)$. As $\tau$ decreases, these move towards $s=1/2$ where they collide for $\aw=\alpha$. As $\tau$ decreases further, the stationary points acquire an opposite nonvanishing imaginary part and move away from the real axis.

When evaluated on the critical points, the second derivative of $\phi(s)$ reads
\begin{align}
    \phi''(s_\pm) &= 
    \begin{cases}
        \displaystyle\mp8\phantom{\mathrm{i}}\frac{\alpha^2}{\aw^2}\sqrt{1-\frac{\aw}{\alpha}} \qquad \text{for $\alpha>\aw$ ,} \\[1em]
        \displaystyle\mp8\mathrm{i}\frac{\alpha^2}{\aw^2}\sqrt{\frac{\aw}{\alpha}-1} \qquad \text{for $\alpha<\aw$ .}
    \end{cases}
\end{align}
When $\alpha>\aw$ and both $s_{\pm}$ sit on the real axis, $\phi''(s_+)<0$, while $\phi''(s_-)>0$. In other words, when applying the Laplace approximation method, we only consider the contribution coming from $s_-$, which is a minimum for $\phi(s)$ and corresponds to the perturbative saddle. On the other hand, the contribution coming from the nonperturbative saddle $s_+$ is suppressed. We already know what the large-$N$ asymptotics in this regime is, as it is the result computed in Section~\ref{SEC:large_N}.

The coalescence of the two saddle points is responsible for a critical behavior: when $\alpha<\aw$, the nonperturbative saddle is no longer suppressed and needs to be considered. The function $\phi(s)$ has the same real part when evaluated on both saddles. Specifically,
\begin{align}
    \phi(s_\pm) = \frac{\alpha}{\aw}\left(1\pm\mathrm{i}\sqrt{\frac{\aw}{\alpha}-1}\right) \mp \mathrm{i}\arctan\mleft(\sqrt{\frac{\aw}{\alpha}-1}\mright) \pm \mathrm{i}\frac{\pi}{2} \;.
\end{align}
The integral can be conveniently computed by deforming the original contour $\gamma$ into a $\gamma'$ that traverses the saddles along the associated steepest-descent path. As shown in Figure~\ref{FIG:contours}, $\gamma'$ crosses the saddle in $s_+$ with $\arg{s} = \pi/4$ and the saddle in $s_-$ with $\arg{s} = 3\pi/4$. This gives the large-$N$ asymptotics
\begin{align}
    \mathpzc{z}_{\boldsymbol{0}}(\alpha,\tau)
    \sim {}&\frac{e^{5N^2/4}}{(\aw/\alpha-1)^{1/4}} \left(-\frac{12+\tau}{12\tau}\right)^{N^2/2} \exp\!\bigg(\frac{N^2\alpha(6+\tau)}{2\tau(12+\tau)}\bigg) \cr
    &\times\cos\mleft(N^2\left[\frac{\alpha}{\aw}\sqrt{\frac{\aw}{\alpha}-1} - \arctan\mleft(\sqrt{\frac{\aw}{\alpha}-1}\mright)\right] - \frac{\pi}{4}\mright) \;.
\end{align}
We observe that the above expression does not have a definite sign. In fact, it oscillates rapidly when $N$ is large: it is clear that the full theory cannot be dominated by just the zero-flux sector for $\alpha<\aw$. This regime is thus characterized by the presence of nonperturbative terms both in the effective 't Hooft coupling $\alpha$ and in the rescaled deformation parameter $\tau$. We will denote this region of the phase diagram as the \emph{mixed phase}.\footnote{
    The phase diagram at $\tau<0$ and the role of the nonperturbative saddles was studied in \cite{Gorsky:2020qge}, although the results therein do not quite agree with our findings.
}

We can say more: when the system is in the weak-coupling phase and approaches the critical line at $\alpha=\alpha_{\mathrm{c}}$, we argue that it exhibits a behavior typical of a system in an ordered phase approaching a second-order phase transition. To this aim, we loosely identify $\alpha$ as an inverse temperature in the following. We can compute the ``specific heat''
\begin{align}
    C
    &= \alpha^2\partial^2_\alpha F_0(\alpha,\tau) \cr
    &= \frac{1}{2\sqrt{1-\alpha_{\mathrm{c}}/\alpha}} \;,
\end{align}
to find that the associated critical exponent is $1/2$.

This behavior is not specific to the weak-coupling phase, but rather it is typical of any envelope of characteristics of the Burgers equation \eqref{EQ:f_equation}. In fact, if we write $C$ in the language of Section~\ref{SEC:large_N}, we find
\begin{align}
    C
    &= \frac{\y}{4} \Big(\y\,\partial_{\y} \mathcal{E}(\y,\tau) - \mathcal{E}(\y,\tau)\Big) \cr
    &= \frac{\y}{4} \Big(\y\,\dot{\varphi}(\xi(\y,\tau))\,\partial_{\y}\xi(\y,\tau) - \varphi(\xi(\y,\tau)) \Big)
\end{align}
As mentioned in Section~\ref{SEC:large_N}, the condition leading to an envelope of characteristics is that
\begin{align}\label{EQ:envelope_condition}
    \frac{1}{\partial_{\y}\xi(\y,\tau)} &= 1+\tau\,\dot{\varphi}(\xi(\y,\tau))
\end{align}
should vanish as $\y$ approaches the critical value $\y_{\text{c}}$. As a consequence, on every envelope of characteristics, $C$ diverges.
Although its derivative is singular, $\xi$ is finite on the envelope. We are therefore led to the ansatz
\begin{align}
    \xi(\y,\tau) &= \xi(\y_{\text{c}},\tau) + \xi_1(\tau)\,(\y-\y_{\text{c}})^{\gamma} + \ldots
\end{align}
where $0<\gamma<1$ and the dots represent subleading terms. Now we use \eqref{EQ:envelope_condition} to fix the leading power in $\y-\y_{\text{c}}$ and the associated coefficient. In particular, from
\begin{align}
    \frac{(\y-\y_{\text{c}})^{1-\gamma}}{\gamma\,\xi_1(\tau)} + \ldots
    &= \tau\,\ddot{\varphi}(\xi(\y_{\text{c}},\tau)) \, \xi_1(\tau)\,(\y-\y_{\text{c}})^{\gamma} + \ldots \;,
\end{align}
we deduce that
\begin{align}
    \partial_{\y}\xi(\y,\tau) &\sim \frac{(\y-\y_{\text{c}})^{-1/2}}{\sqrt{2\tau\,\ddot{\varphi}(\xi(\y_{\text{c}},\tau))}} \;.
\end{align}
This, in turn, leads to
\begin{align}
    C
    &\sim -\frac{\y_{\text{c}}^2}{4\tau\sqrt{2\tau\,\ddot{\varphi}(\xi(\y_{\text{c}},\tau))}} \, (\y-\y_{\text{c}})^{-1/2} \;,
\end{align}
which, once again, gives $C \sim (\alpha-\alpha_{\text{c}})^{-1/2}$.
What we have just proven can be checked to reproduce exactly the result in the weak-coupling phase if we identify $\aw^{\vphantom{2}}=\y_{\text{c}}^2$, but it applies also to the envelope associated with the strong-coupling  characteristics upon identifying $\alpha_{\text{s}}^{\vphantom{2}}=\y_{\text{c}}^2$.\footnote{
    We can either regard $\alpha_{\text{s}}$ as a function of $\tau$ or assume that both $\y_{\text{c}}$ and $\tau$ are functions of $k$ as in \eqref{EQ:strong_envelope}.
}
We conclude that the black envelope line in Figure~\ref{FIG:characteristics} at the boundary of the mixed phase can thus be thought of as a single continuous critical line with associated a second order phase transition with critical exponent $1/2$.

\section*{Acknowledgments}
The research of R.P.\ is partly supported by the Knut and Alice Wallenberg Foundation under grant Dnr KAW 2015.0083.

\bibliographystyle{JHEP}
\bibliography{bibliography}
\end{document}